\documentstyle[12pt]{article}
\begin{document}
\title{Non-Noether symmetries in integrable models}
\author{George Chavchanidze} \date{} \maketitle
\thanks{Department of Theoretical Physics, A. Razmadze Institute of Mathematics, 1 Aleksidze Street, Tbilisi 0193, Georgia}
\begin{abstract}{\bf Abstract.} In the present paper the non-Noether symmetries of the Toda model, nonlinear Sch\"{o}dinger equation and Korteweg-de Vries equations (KdV and mKdV) are discussed. It appears that these symmetries yield the complete sets of conservation laws in involution and lead to the bi-Hamiltonian realizations of the above mentioned models.\end{abstract}
Because of their exceptional properties the non-Noether symmetries could be effectively used in analysis of Hamiltonian dynamical systems. From the geometric point of view these symmetries are important because of their tight relationship with geometric structures on phase space such as bi-Hamiltonian structures, Fr\"{o}licher-Nijenhuis operators, Lax pairs and bicomplexes \cite{r1}. The correspondence between non-Noether symmetries and conservation laws is also interesting and in regular Hamiltonian systems on $2n$ dimensional Poisson manifold up to $n$ integrals of motion could be associated with each generator of non-Noether symmetry \cite{r1} \cite{r3}. As a result non-Noether symmetries could be especially useful in analysis of Hamiltonian systems with many degrees of freedom, as well as infinite dimensional Hamiltonian systems, where large (and even infinite) number of conservation laws could be constructed from the single generator of such a symmetry. Under certain conditions satisfied by the symmetry generator these conservation laws appear to be involutive and ensure integrability of the dynamical system. \\
The n-particle non periodic Toda model is one of integrable models that possesses such a nontrivial symmetry. In this model non-Noether symmetry (which is one-parameter group of noncannonical transformations) yields conservation laws that appear to be functionally independent, involutive and ensure the integrability of this dynamical system. Well known bi-Hamiltonian realization of the Toda model is also related to this symmetry. \\
Nonlinear Schr\"{o}dinger equation is another important example where symmetry (again one-parameter group) leads to the infinite sequence of conservation laws in involution. The KdV and mKdV equations also possess non-Noether symmetries which are quite nontrivial (but symmetry group is still one-parameter) and in each model the infinite set of conservation laws is associated with the single generator of the symmetry. \\
Before we consider these models in detail we briefly remind some basic facts concerning symmetries of Hamiltonian systems. Since throughout the article continuous one-parameter groups of symmetries play central role let us remind that each vector field $E$ on the phase space $M$ of the Hamiltonian dynamical system defines continuous one-parameter group of transformations (flow) 
\begin{eqnarray}
g_{a} = e^{aL_{E}}
\end{eqnarray}
where $L_{E}$ denotes Lie derivative along the vector field $E$. Action of this group on observables (smooth functions on $M$) is given by expansion 
\begin{eqnarray}
g_{a}(f) = e^{aL_{E}}(f) = f + aL_{E}f + \frac{1}{2}a^{2}L_{E}^{2}f + ...
\end{eqnarray}
Further it will be assumed that $M$ is $2n$ dimensional symplectic manifold and the group of transformations $g_{a}$ will be called symmetry of Hamiltonian system if it preserves manifold of solutions of Hamilton's equation 
\begin{eqnarray} \label{eq:e1}
\frac{ d }{dt }f = \{h , f\}
\end{eqnarray}
(here $\{ , \}$ denotes Poisson bracket defined in a standard manner by Poisson bivector field $\{f , g\} = W(df \wedge dg)$ and $h$ is smooth function on $M$ called Hamiltonian) or in other words if for each $f$ satisfying Hamilton's equation $g_{a}(f)$ also satisfies it. This happens when $g_{a}$ commutes with time evolution operator 
\begin{eqnarray}
\frac{ d }{dt }g_{a}(f) = g_{a}(\frac{ d }{dt }f) 
\end{eqnarray}
If in addition the generator $E$ of the group $g_{a}$ does not preserve Poisson bracket structure $[E , W] \neq 0$ then the $g_{a}$ is called non-Noether symmetry. Let us briefly recall some basic features of non-Noether symmetries. First of all if $E$ generates non-Noether symmetry then the $n$ functions 
\begin{eqnarray} \label{eq:e2}
Y_{k} = i_{W^{k}}(L_{E}\omega )^{k} ~~~~~k = 1,2, ... n
\end{eqnarray}
(where $\omega $ is symplectic form obtained by inverting Poisson bivector $W$ and $i$ denotes contraction) are integrals of motion (see \cite{r1} \cite{r3}) and if additionally the symmetry generator $E$ satisfies Yang-Baxter equation 
\begin{eqnarray} \label{eq:e3}
[[E[E , W]]W] = 0
\end{eqnarray}
these conservation laws $Y_{k}$ appear to be in involution $\{Y_{k}, Y_{m}\} = 0$ while the bivector fields $W$ and $[E , W]$ (or in terms of 2-forms $\omega $ and $L_{E}\omega $) form bi-Hamiltonian system (see \cite{r1}). Due to this features non-Noether symmetries could be effectively used in construction of conservation laws and bi-Hamiltonian structures. \\
Now let us focus on non-Noether symmetry of the Toda model -- $2n$ dimensional Hamiltonian system that describes the motion of $n$ particles on the line governed by the exponential interaction. Equations of motion of the non periodic n-particle Toda model are 
\begin{eqnarray} \label{eq:e4}
\frac{ d }{dt }q_{i} = p_{i}\nonumber \\\frac{ d }{dt }p_{i} = \epsilon (i - 1)e^{q_{i - 1} - q_{i}} - \epsilon (n - i)e^{q_{i} - q_{i + 1}} 
\end{eqnarray}
($\epsilon (k) = - \epsilon (- k) = 1$ for any natural $k$ and $\epsilon (0) = 0$) and could be rewritten in Hamiltonian form (\ref{eq:e1}) with canonical Poisson bracket defined by 
\begin{eqnarray}
W = \sum ^{ n }_{i = 1 }\frac{ \partial }{\partial p_{i} }\wedge \frac{ \partial }{\partial q_{i} }
\end{eqnarray}
corresponding symplectic form 
\begin{eqnarray}
\omega = \sum ^{ n }_{i = 1 }dp_{i} \wedge dq_{i} 
\end{eqnarray}
and Hamiltonian equal to 
\begin{eqnarray}
h = \frac{ 1 }{2 }\sum ^{ n }_{i = 1 }p_{i}^{2} + \sum ^{ n - 1 }_{i = 1 }e^{q_{i} - q_{i + 1}} 
\end{eqnarray}
The group of transformations $g_{a}$ generated by the vector field $E$ will be symmetry of Toda chain if for each $p_{i}, q_{i}$ satisfying Toda equations (\ref{eq:e4}) $g_{a}(p_{i}), g_{a}(q_{i})$ also satisfy it. Substituting infinitesimal transformations 
\begin{eqnarray}
g_{a}(p_{i}) = p_{i} + aE(p_{i}) + O(a^{2})\nonumber \\g_{a}(p_{i}) = q_{i} + aE(q_{i}) + O(a^{2}) 
\end{eqnarray}
into (\ref{eq:e4}) and grouping first order terms gives rise to the conditions 
\begin{eqnarray} \label{eq:e5}
\frac{ d }{dt }E(q_{i}) = E(p_{i})\nonumber \\\frac{ d }{dt }E(p_{i}) = \epsilon (i - 1)e^{q_{i - 1} - q_{i}} (E(q_{i - 1}) - E(q_{i})) - \epsilon (n - i)e^{q_{i} - q_{i + 1}} (E(q_{i}) - E(q_{i + 1})) 
\end{eqnarray}
One can verify that the vector field defined by 
\begin{eqnarray} \label{eq:e6}
E(p_{i}) = \frac{ 1 }{2 }p_{i}^{2} + \epsilon (i - 1)(n - i + 2) e^{q_{i - 1} - q_{i}} - \epsilon (n - i)(n - i) e^{q_{i} - q_{i + 1}} +\nonumber \\\frac{ t }{2 }(\epsilon (i - 1)(p_{i - 1} + p_{i}) e^{q_{i - 1} - q_{i}} - \epsilon (n - i)(p_{i} + p_{i + 1}) e^{q_{i} - q_{i + 1}} \nonumber \\E(q_{i}) = (n - i + 1)p_{i} - \frac{ 1 }{2 }\sum ^{ i - 1 }_{k = 1 }p_{k} + \frac{ 1 }{2 }\sum ^{ n }_{k = i + 1 }p_{k} + \nonumber \\\frac{ t }{2 }(p_{i}^{2} + \epsilon (i - 1)e^{q_{i - 1} - q_{i}} + \epsilon (n - i)e^{q_{i} - q_{i + 1}}) 
\end{eqnarray}
satisfies (\ref{eq:e5}) and generates symmetry of Toda chain. It appears that this symmetry is non-Noether since it does not preserve Poisson bracket structure $[E , W] \neq 0$ and additionally one can check that Yang-Baxter equation $[[E[E , W]]W] = 0$ is satisfied. This symmetry could play important role in analysis of Toda model. First let us note that calculating $L_{E}\omega $ leads to the following 2-form 
\begin{eqnarray}
L_{E}\omega = \sum ^{ n }_{i = 1 }p_{i}dp_{i} \wedge dq_{i} + \sum ^{ n - 1 }_{i = 1 }e^{q_{i} - q_{i + 1}} dq_{i} \wedge q_{i + 1} + \sum ^{ ~ }_{i < j }dp_{i} \wedge dp_{j} 
\end{eqnarray}
and together $\omega $ and $L_{E}\omega $ give rise to bi-Hamiltonian structure of Toda model (compare with \cite{r2}). The conservation laws (\ref{eq:e2}) associated with the symmetry reproduce well known set of conservation laws of Toda chain. 
\begin{eqnarray}
I_{1} = Y_{1} = \sum ^{ n }_{i = 1 }p_{1} + p_{2}\nonumber \\I_{2} = \frac{ 1 }{2 }Y_{1}^{2} - Y_{2} = \frac{ 1 }{2 }\sum ^{ n }_{i = 1 }p_{i}^{2} + \sum ^{ n - 1 }_{i = 1 }e^{q_{i} - q_{i + 1}}\nonumber \\I_{3} = \frac{ 1 }{3 }Y_{1}^{3} - Y_{1}Y_{2} + Y_{3} = \frac{ 1 }{3 }\sum ^{ n }_{i = 1 }p_{i}^{3} + \sum ^{ n - 1 }_{i = 1 }(p_{i} + p_{i + 1}) e^{q_{i} - q_{i + 1}}\nonumber \\I_{4} = \frac{ 1 }{4 }Y_{1}^{4} - Y_{1}^{2}Y_{2} + \frac{ 1 }{2 }Y_{2}^{2} + Y_{1}Y_{3} - Y_{4} = \nonumber \\\frac{ 1 }{4 }\sum ^{ n }_{i = 1 }p_{i}^{4} + \sum ^{ n - 1 }_{i = 1 }(p_{i}^{2} + 2p_{i}p_{i + 1} + p_{i + 1}^{2}) e^{q_{i} - q_{i + 1}} + \frac{ 1 }{2 }\sum ^{ n - 1 }_{i = 1 }e^{2(q_{i} - q_{i + 1})} + \sum ^{ n - 2 }_{i = 1 }e^{q_{i} - q_{i + 2}} \nonumber \\I_{m} = (- 1)^{m}Y_{m} + m^{- 1} \sum ^{ m - 1 }_{k = 1 }(- 1)^{k}I_{m - k}Y_{k} 
\end{eqnarray}
The condition $[[E[E , W]]W] = 0$ satisfied by generator of the symmetry $E$ ensures that the conservation laws are in involution i. e. $\{Y_{k},Y_{m}\} = 0$. Thus the conservation laws as well as the bi-Hamiltonian structure of the non periodic Toda chain appear to be associated with non-Noether symmetry. \\
Unlike the Toda model the dynamical systems in our next examples are infinite dimensional and in order to ensure integrability one should construct infinite number of conservation laws. Fortunately in several integrable models this task could be effectively done by identifying appropriate non-Noether symmetry. First let us consider well known infinite dimensional integrable Hamiltonian system -- nonlinear Schr\"{o}dinger equation (NSE) 
\begin{eqnarray}
\psi _{t} = i(\psi _{xx} + 2\psi ^{2}\bar\psi )
\end{eqnarray}
where $\psi $ is a smooth complex function of $(t, x) \in R^{2}$. On this stage we will not specify any boundary conditions and will just focus on symmetries of NSE. Supposing that the vector field $E$ generates the symmetry of NSE one gets the following restriction 
\begin{eqnarray} \label{eq:e7}
E(\psi )_{t} = i[E(\psi )_{xx} + 2\psi ^{2}E(\bar\psi ) + 4\psi \bar\psi E(\psi )]
\end{eqnarray}
(obtained by substituting infinitesimal transformation $\psi \rightarrow \psi + aE(\psi ) + O(a^{2})$ generated by $E$ into NSE). It appears that NSE possesses nontrivial symmetry that is generated by the vector field 
\begin{eqnarray}
E(\psi ) = i(\psi _{x} + \frac{ x }{2 }\psi _{xx} + \psi \phi + x\psi ^{2}\bar\psi ) - t(\psi _{xxx} + 6\psi \bar\psi \psi _{x})
\end{eqnarray}
(here $\phi $ is defined by $\phi _{x} = \psi \bar\psi $). In order to construct conservation laws we also need to know Poisson bracket structure and it appears that invariant Poisson bivector field could be defined if $\psi $ is subjected to either periodic $\psi (t, - \infty ) = \psi (t, + \infty )$ or zero $\psi (t, - \infty ) = \psi (t, + \infty ) = 0$ boundary conditions. In terms of variational derivatives the explicit form of the Poisson bivector field is 
\begin{eqnarray}
W = i \int ^{ + \infty }_{- \infty }dx \frac{ \delta }{\delta \psi }\wedge \frac{ \delta }{\delta \bar\psi }
\end{eqnarray}
while corresponding symplectic form obtained by inverting $W$ is 
\begin{eqnarray}
\omega = i \int ^{ + \infty }_{- \infty }dx \delta \psi \wedge \delta \bar\psi 
\end{eqnarray}
Now one can check that NSE could be rewritten in Hamiltonian form 
\begin{eqnarray}
\psi _{t} = \{h , \psi \}
\end{eqnarray}
with Poisson bracket $\{ , \}$ defined by $W$ and 
\begin{eqnarray}
h = \int ^{ + \infty }_{- \infty }dx (\psi ^{2}\bar\psi ^{2} - \psi _{x}\bar\psi _{x}) 
\end{eqnarray}
Knowing the symmetry of NSE that appears to be non-Noether ($[E, W] \neq 0$) one can construct bi-Hamiltonian structure and conservation laws. First let us calculate Lie derivative of symplectic form along the symmetry generator 
\begin{eqnarray}
L_{E}\omega = \int ^{ + \infty }_{- \infty }[\delta \psi _{x} \wedge \delta \bar\psi + \psi \delta \phi \wedge \delta \bar\psi + \bar\psi \delta \phi \wedge \delta \psi ]dx 
\end{eqnarray}
The couple of 2-forms $\omega $ and $L_{E}\omega $ exactly reproduces the bi-Hamiltonian structure of NSE proposed by Magri \cite{r4} while the conservation laws associated with this symmetry are well known conservation laws of NSE 
\begin{eqnarray}
I_{1} = Y_{1} = 2 \int ^{ + \infty }_{- \infty }\psi \bar\psi dx\nonumber \\I_{2} = Y_{1}^{2} - 2Y_{2} = i \int ^{ + \infty }_{- \infty }(\bar\psi _{x}\psi - \psi _{x}\bar\psi ) dx\nonumber \\I_{3} = Y_{1}^{3} - 3Y_{1}Y_{2} + 3Y_{3} = 2 \int ^{ + \infty }_{- \infty } (\psi ^{2}\bar\psi ^{2} - \psi _{x}\bar\psi _{x}) dx \nonumber \\I_{4} = Y_{1}^{4} - 4Y_{1}^{2}Y_{2} + 2Y_{2}^{2} + 4Y_{1}Y_{3} - 4Y_{4} = \nonumber \\\int ^{ + \infty }_{- \infty }[i(\bar\psi _{x}\psi _{xx} - \psi _{x}\bar\psi _{xx}) + 3i(\bar\psi \psi ^{2}\bar\psi _{x} - \psi \bar\psi ^{2}\psi _{x})] dx \nonumber \\I_{m} = (- 1)^{m}mY_{m} + \sum ^{ m - 1 }_{k = 1 }(- 1)^{k}I_{m - k}Y_{k} 
\end{eqnarray}
The involutivity of the conservation laws of NSE $\{Y_{k}, Y_{m}\} = 0 $ is related to the fact that $E$ satisfies Yang-Baxter equation $[[E[E , W]]W] = 0$. \\
Now let us consider other important integrable models -- Korteweg-de Vries equation (KdV) and modified Korteweg-de Vries equation (mKdV). Here symmetries are more complicated but generator of the symmetry still can be identified and used in construction of conservation laws. The KdV and mKdV equations have the following form 
\begin{eqnarray}
u_{t} + u_{xxx} + uu_{x} = 0 [KdV]
\end{eqnarray}
and 
\begin{eqnarray}
u_{t} + u_{xxx} - 6u^{2}u_{x} = 0 [mKdV]
\end{eqnarray}
(here $u$ is smooth function of $(t, x) \in R^{2}$). The generators of symmetries of KdV and mKdV should satisfy conditions 
\begin{eqnarray}
E(u)_{t} + E(u)_{xxx} + u_{x}E(u) + uE(u)_{x} = 0 [KdV]
\end{eqnarray}
and 
\begin{eqnarray}
E(u)_{t} + E(u)_{xxx} - 12uu_{x}E(u) - 6u^{2}E(u)_{x} = 0 [mKdV]
\end{eqnarray}
(again this conditions are obtained by substituting infinitesimal transformation $u \rightarrow u + aE(u) + O(a^{2})$ into KdV and mKdV, respectively). Further we will focus on the symmetries generated by the following vector fields 
\begin{eqnarray}
E(u) = \frac{ 1 }{2 }u_{xx} + \frac{ 1 }{6 }u^{2} + \frac{ 1 }{24 }u_{x}v + \frac{ x }{8 }(u_{xxx} + uu_{x}) - \nonumber \\\frac{ t }{16 }(6u_{xxxxx} + 20u_{x}u_{xx} + 10 uu_{xxx} + 5u^{2}u_{x}) [KdV] 
\end{eqnarray}
and 
\begin{eqnarray}
E(u) = - \frac{ 3 }{2 }u_{xx} + 2u^{3} + u_{x}w - \frac{ x }{2 }(u_{xxx} - 6u^{2}u_{x}) -\nonumber \\\frac{ 3t }{2 }(u_{xxxxx} - 10u^{2}u_{xxx} - 40uu_{x}u_{xx} - 10u_{x}^{3} + 30u^{4}u_{x}) [mKdV]
\end{eqnarray}
(here $v$ and $w$ are defined by $v_{x} = u$ and $w_{x} = u^{2}$) To construct conservation laws we need to know Poisson bracket structure and again like in the case of NSE the Poisson bivector field is well defined when $u$ is subjected to either periodic $u(t, - \infty ) = u(t, + \infty )$ or zero $u(t, - \infty ) = u(t, + \infty ) = 0$ boundary conditions. For both KdV and mKdV the Poisson bivector field is 
\begin{eqnarray}
W = \int ^{ + \infty }_{- \infty }dx \frac{ \delta }{\delta u }\wedge \frac{ \delta }{\delta v }
\end{eqnarray}
with corresponding symplectic form 
\begin{eqnarray}
\omega = \int ^{ + \infty }_{- \infty }dx \delta u \wedge \delta v 
\end{eqnarray}
leading to Hamiltonian realization of KdV and mKdV equations 
\begin{eqnarray}
u_{t} = \{h , u\}
\end{eqnarray}
with Hamiltonians 
\begin{eqnarray}
h = \int ^{ + \infty }_{- \infty }(u_{x}^{2} - \frac{ u^{3} }{3 }) dx [KdV]
\end{eqnarray}
and 
\begin{eqnarray}
h = \int ^{ + \infty }_{- \infty }(u_{x}^{2} + u^{4}) dx [mKdV]
\end{eqnarray}
By taking Lie derivative of the symplectic form along the generators of the symmetries one gets another couple of symplectic forms 
\begin{eqnarray}
L_{E}\omega = \int ^{ + \infty }_{- \infty }dx (\delta u \wedge \delta u_{x} + \frac{ 2 }{3 }u\delta u \wedge \delta v) [KdV] 
\end{eqnarray}
\begin{eqnarray}
L_{E}\omega = \int ^{ + \infty }_{- \infty }dx (\delta u \wedge \delta u_{x} - 2u\delta u \wedge \delta w) [mKdV] 
\end{eqnarray}
involved in bi-Hamiltonian realization of KdV/mKdV hierarchies and proposed by Magri \cite{r4}. The conservation laws associated with the symmetries reproduce infinite sequence of conservation laws of KdV equation 
\begin{eqnarray}
I_{1} = Y_{1} = \frac{ 2 }{3 }\int ^{ + \infty }_{- \infty }u dx \nonumber \\I_{2} = Y_{1} - 2Y_{2} = \frac{ 4 }{9 }\int ^{ + \infty }_{- \infty }u^{2} dx \nonumber \\I_{3} = Y_{1}^{3} - 3Y_{1}Y_{2} + 3Y_{3} = \frac{ 8 }{9 }\int ^{ + \infty }_{- \infty }(\frac{ u^{3} }{3 } - u_{x}^{2}) dx \nonumber \\I_{4} = Y_{1}^{4} - 4Y_{1}^{2}Y_{2} + 2Y_{2}^{2} + 4Y_{1}Y_{3} - 4Y_{4} = \nonumber \\\frac{ 64 }{45 }\int ^{ + \infty }_{- \infty }(\frac{ 5 }{36 }u^{4} - \frac{ 5 }{3 }uu_{x}^{2} + u_{xx}^{2}) dx \nonumber \\I_{m} = (- 1)^{m}mY_{m} + \sum ^{ m - 1 }_{k = 1 }(- 1)^{k}I_{m - k}Y_{k} 
\end{eqnarray}
and mKdV equation 
\begin{eqnarray}
I_{1} = Y_{1} = - 4 \int ^{ + \infty }_{- \infty }u^{2} dx \nonumber \\I_{2} = Y_{1} - 2Y_{2} = 16 \int ^{ + \infty }_{- \infty }(u^{4} + u_{x}^{2}) dx \nonumber \\I_{3} = Y_{1}^{3} - 3Y_{1}Y_{2} + 3Y_{3} = - 32 \int ^{ + \infty }_{- \infty }(2u^{6} + 10 u^{2}u_{x}^{2} + u_{xx}^{2}) dx \nonumber \\I_{4} = Y_{1}^{4} - 4Y_{1}^{2}Y_{2} + 2Y_{2}^{2} + 4Y_{1}Y_{3} - 4Y_{4} = \nonumber \\\frac{ 256 }{5 }\int ^{ + \infty }_{- \infty }(5 u^{8} + 70u^{4}u_{x}^{2} - 7u_{x}^{4} + 14u^{2}u_{xx}^{2} + u_{xxx}^{2}) dx \nonumber \\I_{m} = (- 1)^{m}mY_{m} + \sum ^{ m - 1 }_{k = 1 }(- 1)^{k}I_{m - k}Y_{k} 
\end{eqnarray}
The involutivity of these conservation laws is well known and in terms of the symmetry generators it is ensured by conditions $[[E[E , W]]W] = 0$. Thus the conservation laws and bi-Hamiltonian structures of KdV and mKdV hierarchies are related to the non-Noether symmetries of KdV and mKdV equations. \\
The purpose of the present paper was to illustrate some features of non-Noether symmetries discussed in \cite{r1} and to show that in several important integrable models existence of complete sets of conservation laws could be related to the such symmetries. \\
{\bf Acknowledgements.}
Author is grateful to Abdus Salam International Centre for Theoretical Physics, where essential part
of this article was prepared, for kind hospitality.

\end{document}